# Dual-Band Flexible Endfire Filtering Antenna With Conformal Capability for Emergency Communication Applications

Fan Qin, *Member, IEEE*, Runkai Song, *Student Member, IEEE*, Chao Gu, *Member, IEEE*, Wenchi Cheng, *Senior Member, IEEE* and Steven Gao, *Fellow, IEEE*

***Abstract*—In this letter, a single-layer dual-band flexible conformal filtering endfire antenna is presented. The proposed antenna is based on two co-designed folded dipoles (FDs) working at two frequencies, where the lower-frequency FD acts as a reflector for the higher-frequency one. Then, by devising an additional reflector for lower-frequency FD, dual-band endfire radiation is realized. Parasitic strips are deliberately introduced around the FDs to generate electric coupling and magnetic coupling in the two operating bands, resulting in significant filtering performance with four radiation nulls. With flexible structure and single-layer configuration, the antenna design exhibits flexible conformability with cylindrical surfaces of diverse diameters, thereby enabling seamless integration into scalable emergency communication systems. To verify our design concept, an antenna prototype is fabricated and measured. The measured working frequency ranges from 1.37 to 1.45 GHz and 1.89 to 2.07 GHz. Out-of-band radiation suppression more than 11 dB is achieved under different bending radii. The proposed design offers several advantages including dual-band endfire filtering radiation, flexible conformability and low-profile.**

## I. INTRODUCTION

In emergency rescue operations, professional emergency communication equipment is playing an increasingly important role [1]-[4]. A typical scenario of emergency communication in mountain areas between ground devices and unmanned aerial vehicles (UAVs) is shown in Fig. 1. Due to the signal obstruction prevalent in mountain areas, extendable pole integrated with the emergency communication vehicle are frequently employed to elevate antenna heights to ensure ground-to-air signal transmission. Meanwhile, dual-band or multi-band design is preferred due to its enhanced efficiency in mountainous channels. Additionally, the interference generated by various wireless devices underscores the need for antennas with out-of-band radiation suppression to mitigate multi-band interference. Therefore, flexible endfire antenna designs are favored for reliable ground-to-air communication and structural integrity of extendable poles. In this context, antennas demonstrating flexible conformability, dual-band endfire radiation, and filtering ability are highly desirable.

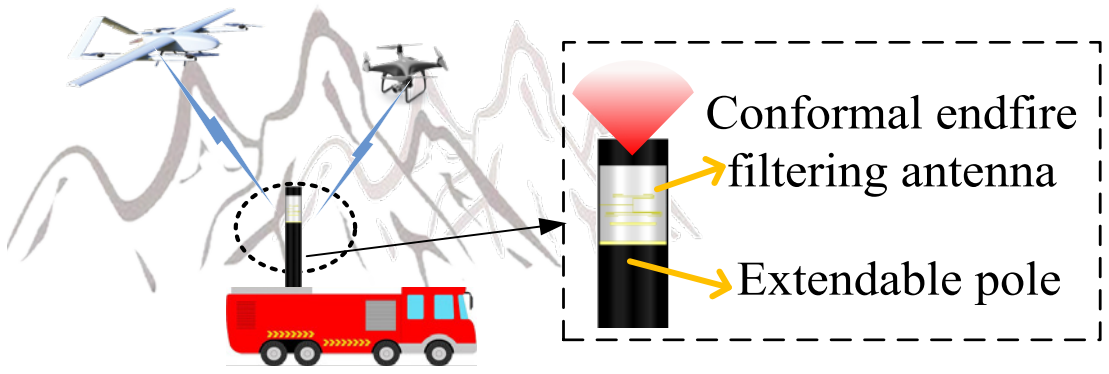

**Fig. 1.** Typical scenario of emergency communication in mountain areas between ground equipment and UAVs.

In recent years, research on filtering antennas has received growing attention [5]-[10]. In [10], a compact omnidirectional filtering antenna is achieved by exciting electric coupling and magnetic coupling on a folded dipole. Moreover, with low insertion loss, compact size and multifunctional capabilities, several designs of dual-band filtering antennas have been proposed [11]-[18]. In [12], a U-slot patch is integrated with dual-mode resonator for dual-band filtering functionality. A dual-band filtering antenna introduced in [14] is achieved by constructing four slots and a microstrip line with open-circuit stepped-impedance resonators. In [15], a single-layer dual-band endfire filtering antenna (EFA) is realized by loading an additional parasitic strip onto a wideband filtering design. A method by inserting vertical metal walls into magnetoelectric dipole for achieving dual band filtering operation is reported in [17]. Although the aforementioned designs offer dual-band filtering characteristics, their applicability in emergency communication scenarios is limited due to a lack of flexible conformability and ease of deployment.

In this letter, a dual-band EFA with flexible conformability is investigated. The design of the dual-band endfire radiation is firstly explained by two connected dipoles with co-designed reflection paths. The introduction of parasitic strips around the connected dipoles results in the formation of four radiation nulls, thereby enhancing the filtering performance. Owing to the newly introduced dual-band EFA structure, the proposed design enables independent tuning of the two frequency bands and the associated four radiation nulls. Furthermore, the compact filtering structure and ultra-low-profile configuration ensure stable filtering performance and robust dual-band

Fan Qin, Runkai Song, and Wenchi Cheng are with the School of Telecommunications Engineering, Xidian University, Xi’an 710071, China (e-mail: fqin@xidian.edu.cn; srk@stu.xidian.edu.cn; wccheng@xidian.edu.cn).

Chao Gu is with the ECIT Institute, Queen’s University Belfast, BT3 9DT Belfast, U.K. (e-mail: chao.gu@qub.ac.uk).

Steven Gao is with the Department of Electronic Engineering, Chinese University of Hong Kong, Hong Kong (e-mail: scgao@ee.cuhk.edu.hk).

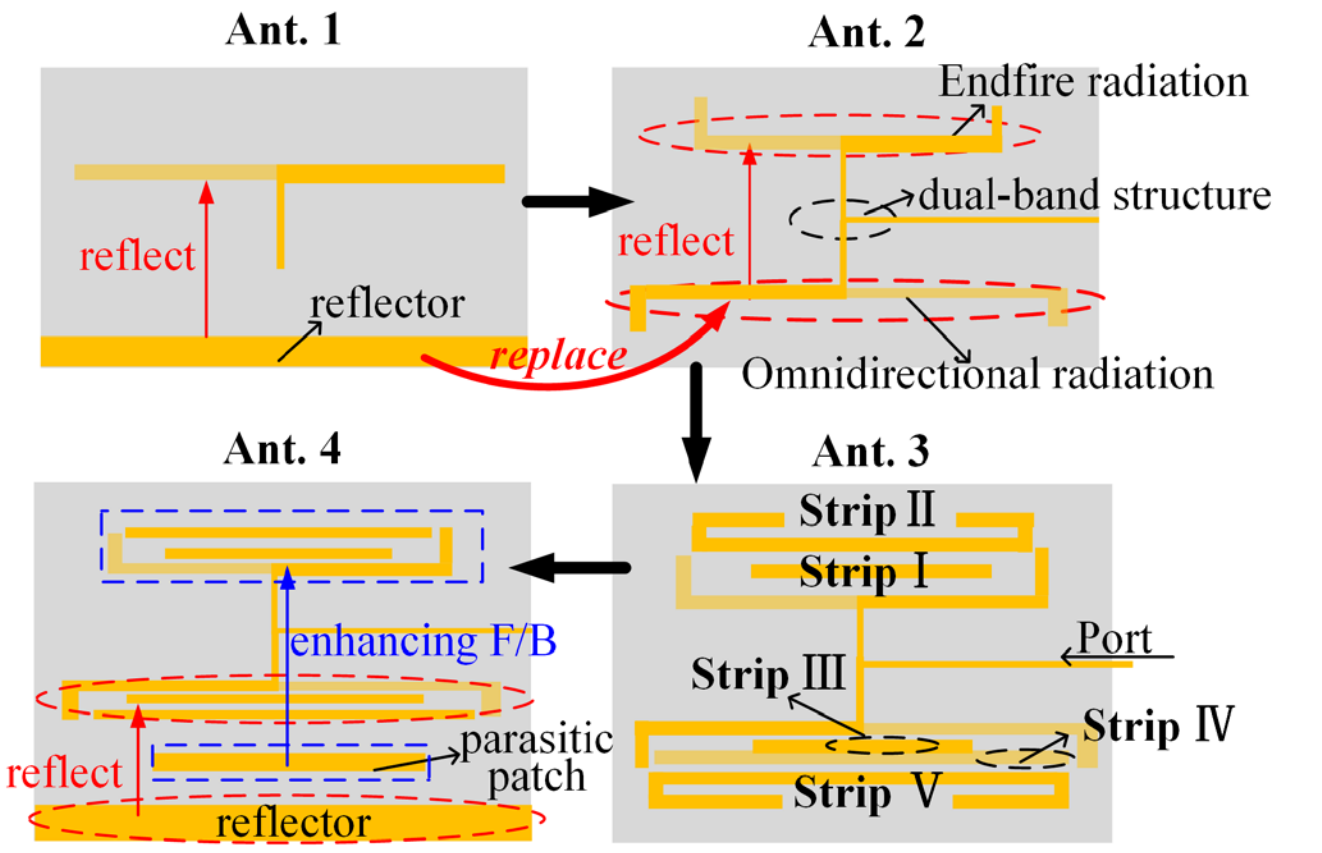

**Fig. 2.** Design procedure of the proposed dual-band endfire filtering antenna.

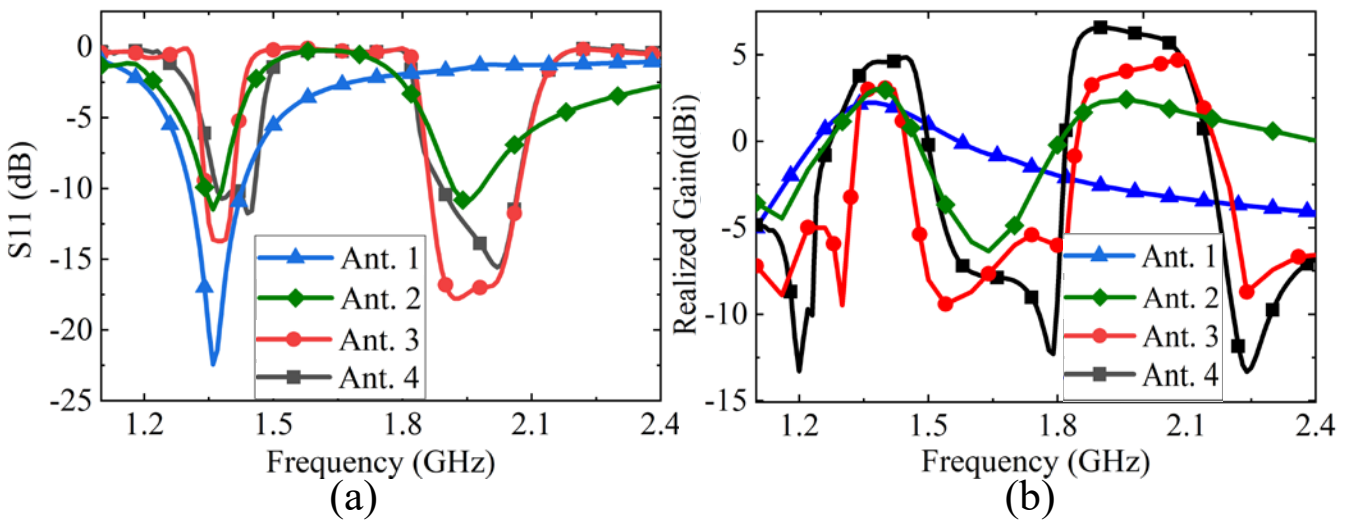

**Fig. 3.** Corresponding $S_{11}$ and realized gain results for Ant. 1 to Ant. 4 shown in Fig. 2.

endfire radiation patterns, even under conformal deformation.

## II. Antenna Design And Analysis

### *A. Design Procedure and Working Principle*

Fig. 2 shows the design procedure of the proposed dual-band flexible conformal EFA. The corresponding $S_{11}$ and realized gain results for Ant. 1 to Ant. 4 are shown in Fig. 3. The design procedures from Ant. 1 to Ant. 4 are as follows.

1) Ant. 1: Baseline structure. The antenna has single-band endfire radiation with a reflector and a driven dipole.

2) Ant. 2: Dual-band configuration. Two dipoles working at different frequencies are connected with T-shaped microstrip line of uniform width, realizing dual-band operation as shown in Fig. 3(a). The longer dipole is subtly replaced as reflector for the other dipole, achieving endfire radiation at higher band (H.B). Meanwhile, the longer dipole working at lower band (L.B) remains omnidirectional radiation.

3) Ant. 3: Dual-band filtering structure. Five parasitic strips are attached to Ant. 2, creating out-of-band radiation suppression with four radiation nulls as shown in Fig. 3(b).

4) Ant. 4: Final design. A reflector is added to reflect the omnidirectional pattern of the longer dipole working at L.B for dual-band endfire radiation. Meanwhile, a parasitic patch is attached to enhance the front-to-back ratio (F/B) at H.B.

Fig. 4 illustrates the field distribution of the dual-band dipole in L.B and H.B. As shown in Fig. 4(a), the longer dipole works at 1.4 GHz and the shorter one works at 2 GHz, respectively, which means EM fields are radiated from longer dipole while the shorter one can be regarded as open circuit when the antenna works at 1.4 GHz. A similar frequency-selective open-circuit phenomenon is also observed at 2 GHz, as illustrated in Fig. 4(b). Furthermore, due to the frequency-selective open-circuit characteristics of the dual-band dipole, the operating frequencies of the two bands can be independently tuned by adjusting the lengths of the connected dipoles. In addition to efficient spatial allocation for the dual-band filtering structure, this design also improves isolation between the two bands.

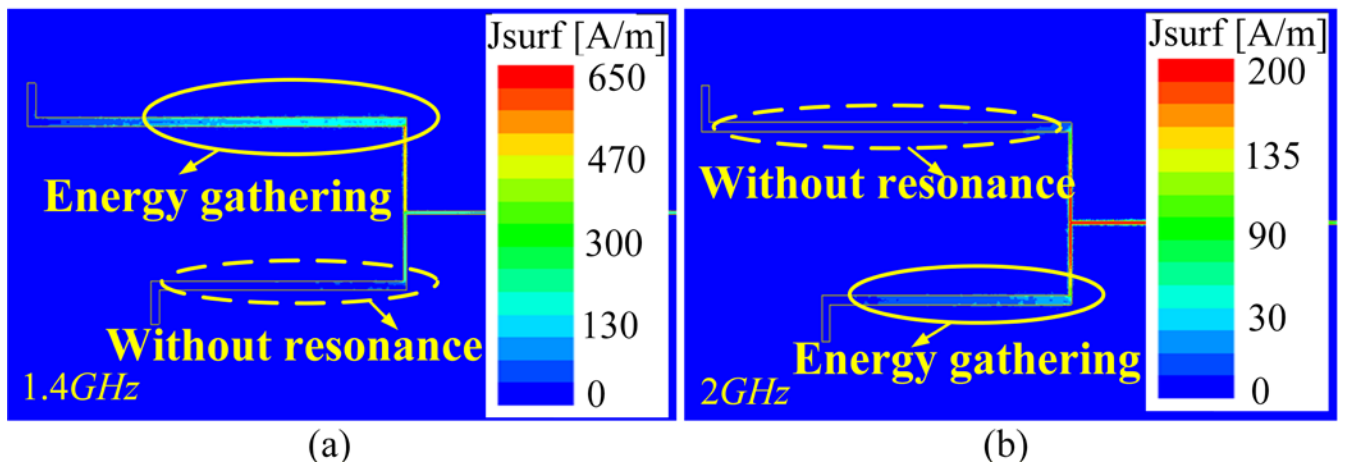

**Fig. 4.** Field distributions of dipole at (a) 1.4GHz, (b) 2GHz.

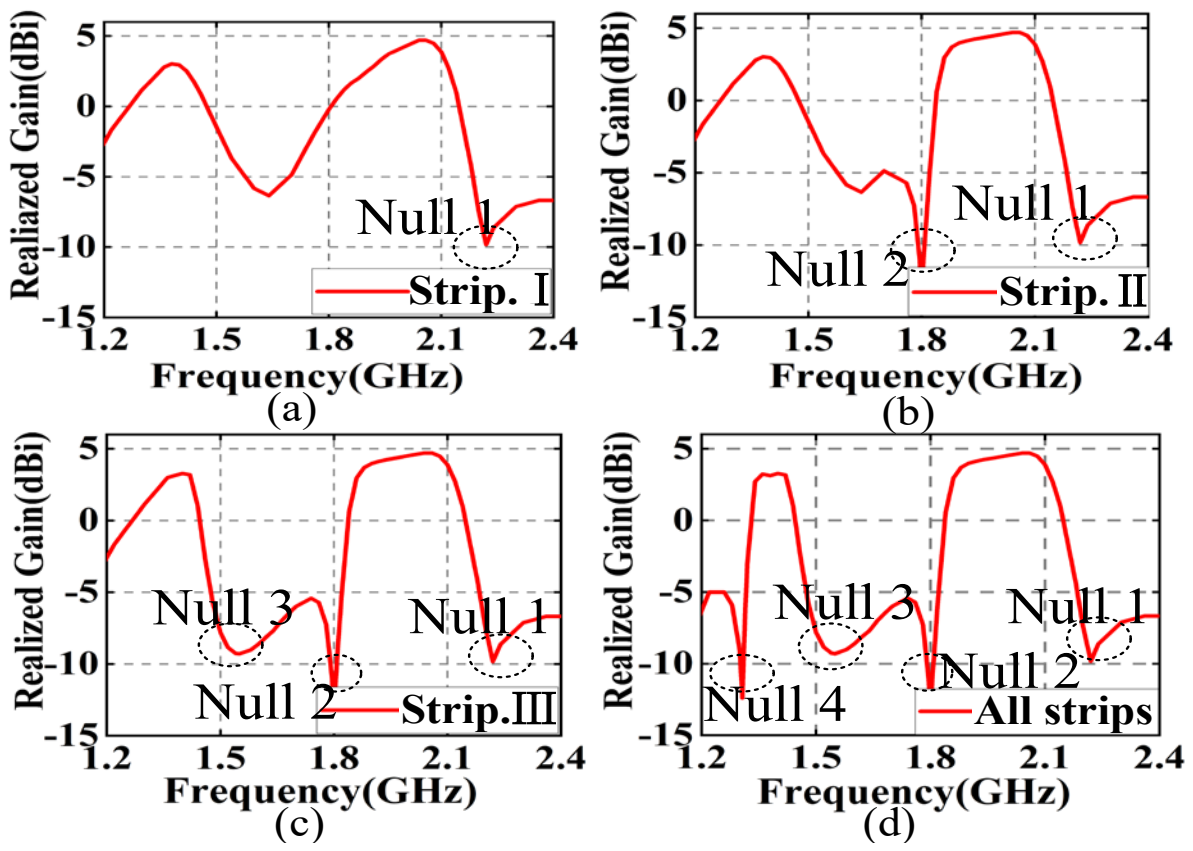

**Fig. 5.** The relationship between radiation nulls and corresponding parasitic strips on Ant. 3.

Due to the magnetic fields are dominant near the center area of the dipole and electric fields are stronger near the end of the dipole, magnetic coupling and electric coupling can be achieved by placing parasitic element in the corresponding location to suppress out-of-band radiation. The relationship between radiation nulls and corresponding parasitic strips on Ant. 3 are shown in Fig. 5. Firstly, strip I is attached near the center of the shorter dipole for electric coupling, generating Null 1 at H.B as shown in Fig. 5(a). Moreover, strip II is loaded on the end of the folded dipole tightly for magnetic coupling, introducing Null 2 at H.B as shown in Fig. 5(b). Similar to the design at H.B, strip III is attached to the longer dipole, creating Null 3 at L.B shown in Fig. 5(c). To achieve both electric and magnetic coupling, strip IV and strip V are printed on both surface of the substrate, achieving Null 4 at L.B as seen in Fig. 5(d).

Fig. 6(a) shows the simulated results before and after attaching reflector. With the structure revolution from Ant. 3 to Ant. 4 shown in Fig. 2, directional diagram of L.B changes from omnidirectional radiation to endfire radiation, with gain in L.B increasing from 3 dBi to 5 dBi. Besides, with adding the parasitic patch, in-band gain of H.B increases from 4 dBi to 6.5 dBi, and the F/B enhances 3 dB shown in Fig. 6(b). Radiation between two passbands is suppressed with a 3 dB reduction. Notably, the out-of-band radiation suppression around 2.4 GHz deteriorates slightly, while the out-of-band suppression of H.B increases due to the raise of in-band gain.

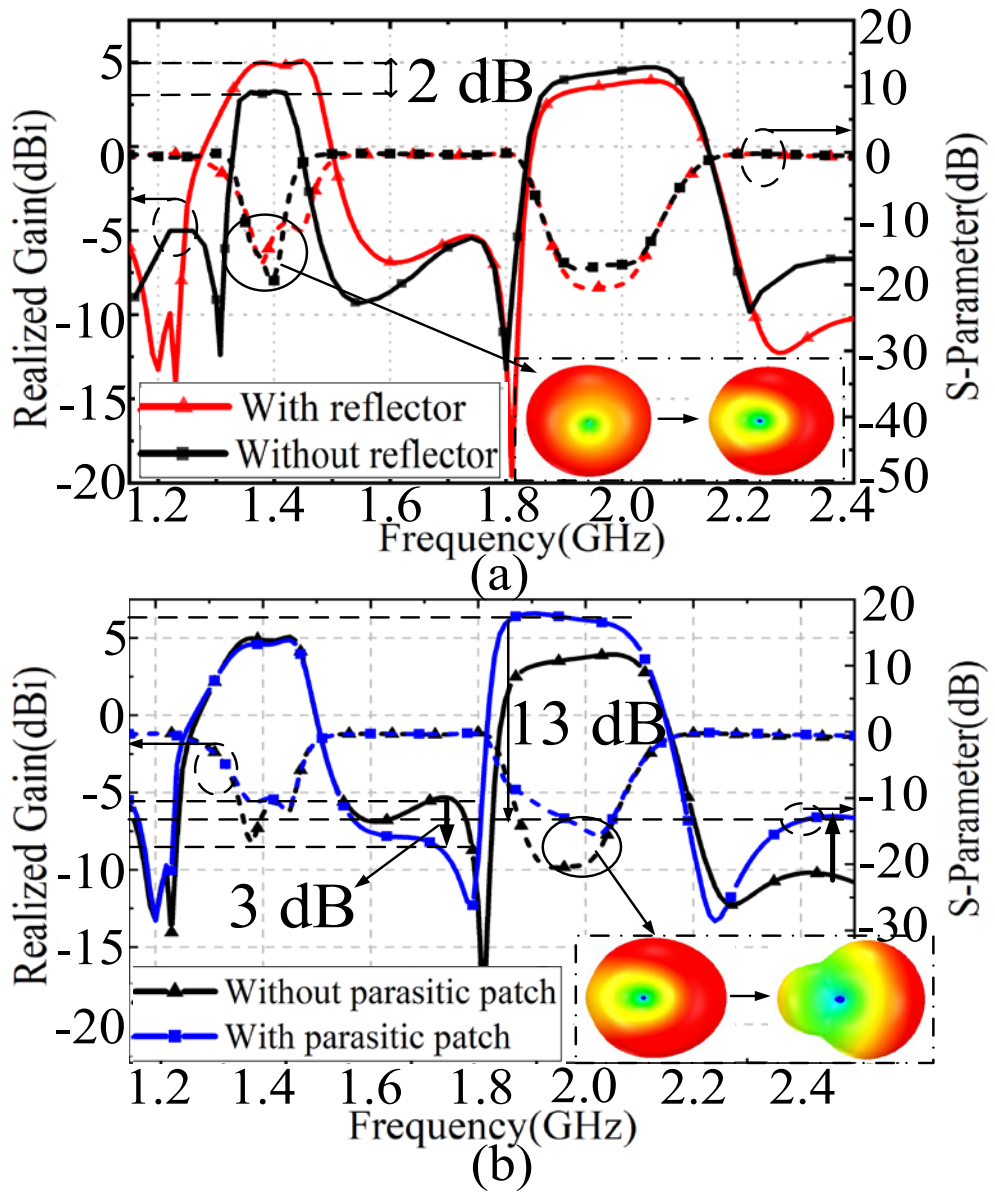

**Fig. 6.** Simulated S11 and gain of the antenna. (a) variation with reflector. (b) variation with parasitic patch.

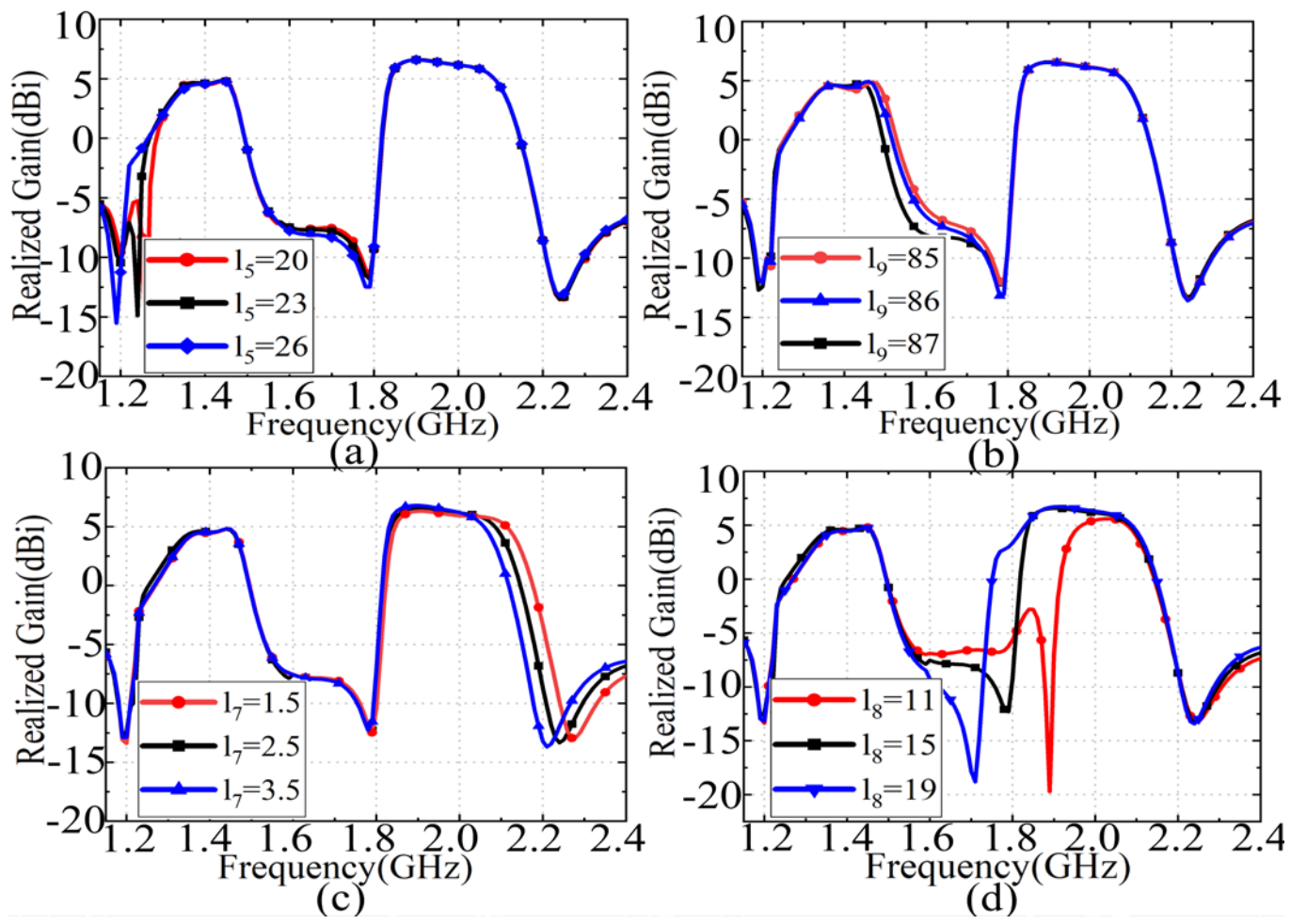

**Fig. 7.** Gain variation with parameters. (a) $l_5$ (b) $l_9$ (c) $l_7$ (d) $l_8$.

*B. Parameter Optimization and Flexible Conformal Analysis*

To address the interaction between the filtering structures in the two frequency bands, parameter optimization of the strips attached to the dual-band dipole is implemented. As for the two parasitic strips added on dipole working at L.B, from Fig. 7(a), we observe that the radiation null in the lower frequency of the L.B shifts from 1.19 to 1.26 GHz when $l_5$ decreases from 26 to 20 mm. Fig. 7(b) reveals that the frequency of radiation null in the higher frequency of the L.B increases from 1.57 to 1.62 GHz when $l_9$ decreases from 87 to 85 mm.

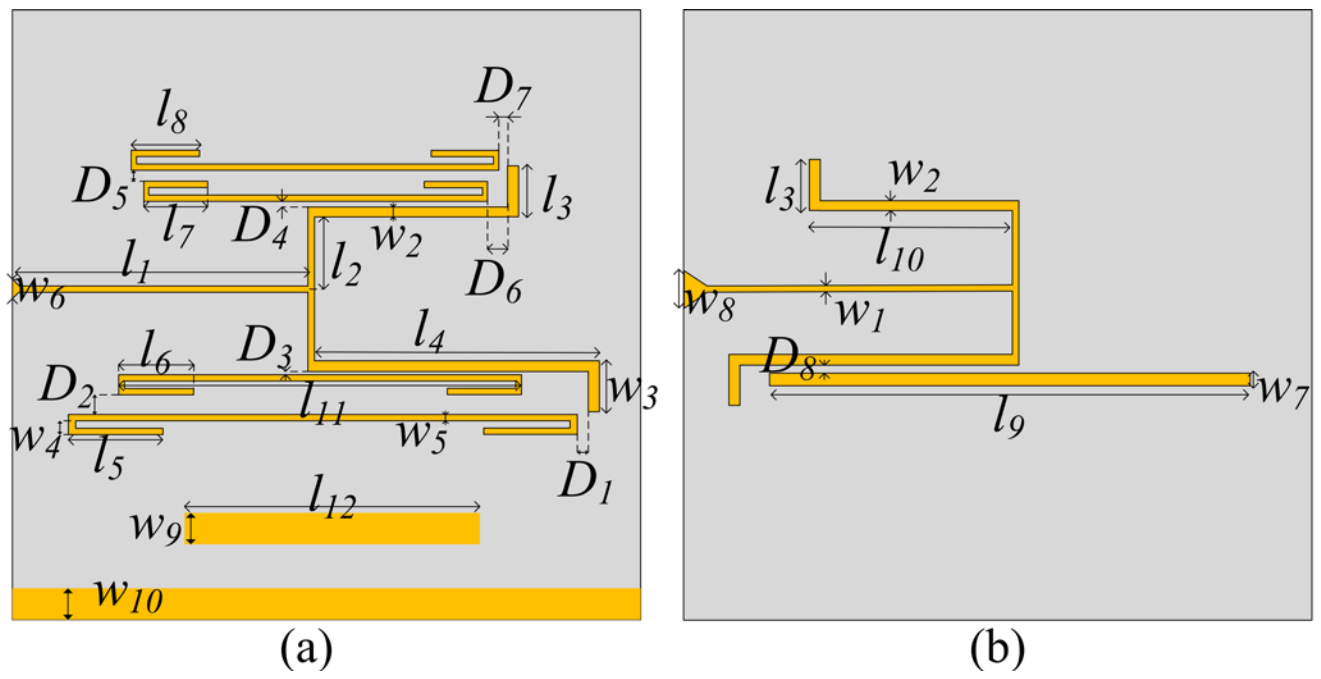

**Fig. 8.** Configuration of the proposed antenna. (a) top view (b) bottom view ($l_1$=60, $l_2$=10, $l_3$=5, $l_4$=45, $l_5$=23, $l_6$=15, $l_7$=2.5, $l_8$=15, $l_9$=87, $l_{10}$=30, $l_{11}$=85, $l_{12}$=69, $w_1$=0.32, $w_2$=1, $w_3$=5, $w_4$=1, $w_5$=0.3, $w_6$=1, $w_7$=1, $w_8$=10, $w_9$=4, $w_{10}$=5, $D_1$=0.3, $D_2$=2.7, $D_3$=0.5, $D_4$=0.8, $D_5$=1.8, $D_6$=1, $D_7$=0.4, $D_8$=0.2, all in mm).

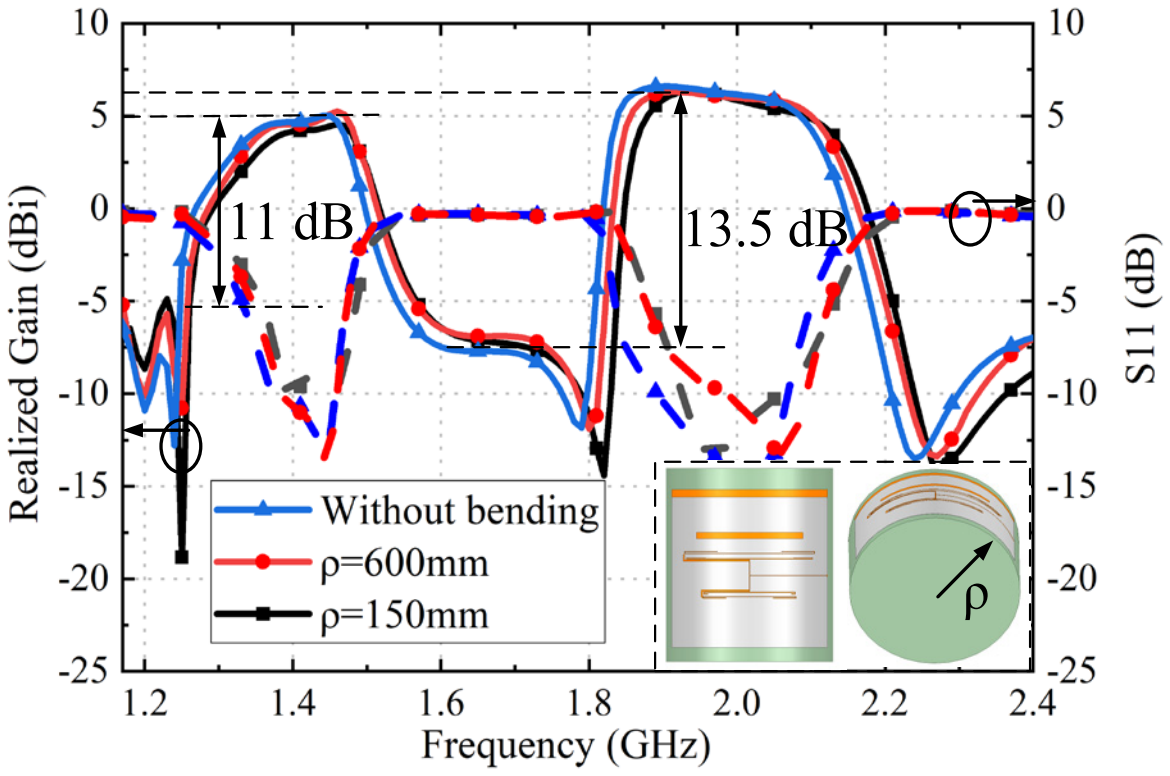

**Fig. 9.** Simulated $S_{11}$ and gain results after conformal to the cylindrical surfaces with different radii ρ.

Moreover, as shown in Fig. 7(c), the radiation null in the higher frequency of the H.B decreases from 2.28 to 2.2 GHz when $l_7$ increases from 1.5 to 3.5 mm. Fig. 7(d) shows that the radiation null in the lower frequency of the H.B increases from 1.7 to 1.9 GHz with $l_8$ decreases from 19 to 11 mm. From the parameter optimization discussed above, it can be seen that the length of parasitic strips can significantly influence the frequencies of the four radiation nulls. After comprehensive optimization, the structure and specific parameters of the proposed antenna are presented in Fig. 8. A flexible F4B substrate with total dimensions of 120 mm × 120 mm × 0.127 mm and a dielectric constant of 2.1 is employed.

To evaluate its performance under bending conditions, the proposed antenna is conformally wrapped around cylindrical surfaces with curvature radii ($\rho$) of 600 mm and 150 mm. As shown in Fig. 9, four radiation nulls keep stable with different bending radius. Meanwhile, the antenna can remain −10 dB impedance bandwidth from 1.37 GHz to 1.45 GHz and 1.89 GHz to 2.07 GHz with great endfire radiation characteristics. Moreover, out-of-band radiation suppression more than 11 dB and 13.5 dB are observed in the L.B and H.B under different bending radius, respectively. The results demonstrate that the antenna maintains stable dual-band endfire filtering performance after flexible conformal deformation.

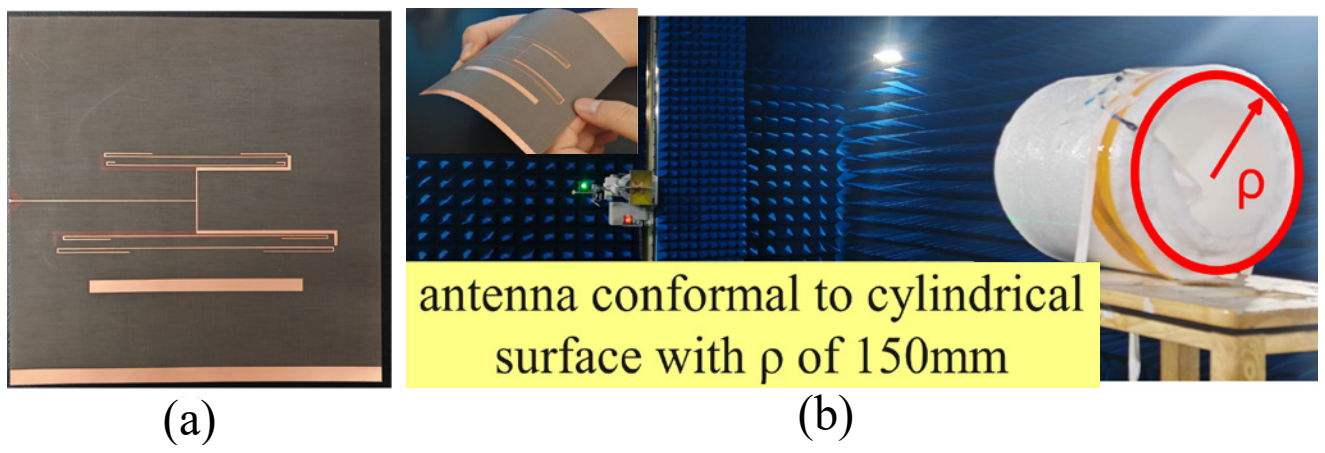

**Fig. 10.** (a) Antenna prototype. (b) Antenna measurement.

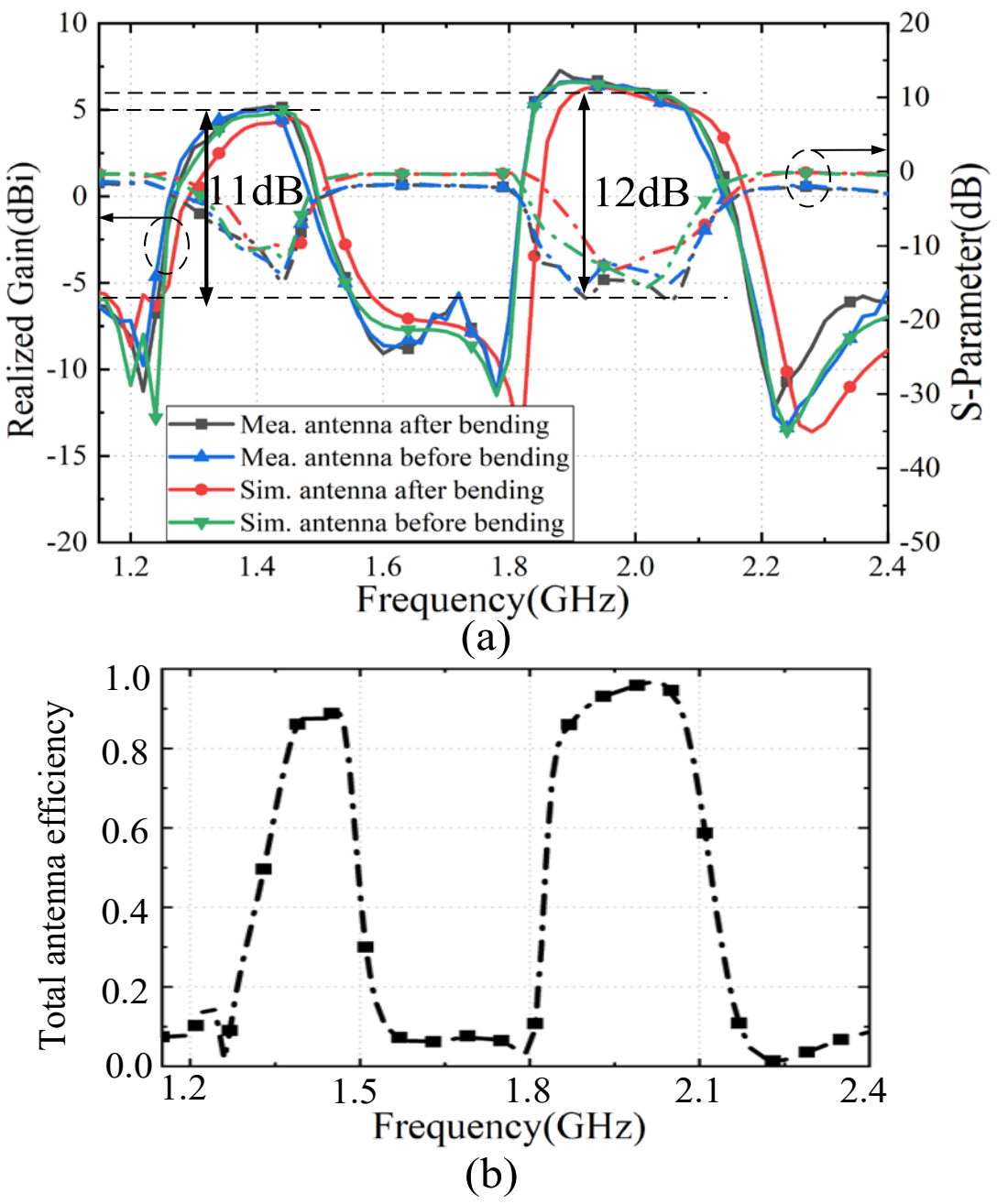

**Fig. 11.** (a) Simulated and measured S11 and gain results. (b) Measured total antenna efficiency (mismatch included).

## III. Antenna Measurement And Discussion

To validate the performance of the proposed antenna, two configurations—planar and conformally wrapped around a foam cylinder with a 150 mm radius—were tested, as shown in Fig. 10. As shown in Fig. 11(a), the measured working frequency is from 1.37 GHz to 1.45 GHz and 1.89 GHz to 2.07 GHz with bandwidth of 5.7% and 9.1%. Additionally, measured in-band gain about 5 dBi in the L.B and 6 dBi in the H.B, out-of-band radiation suppression more than 11 dB in the L.B and 12 dB in the H.B are validated after flexible conformal. As shown in Fig. 11(b), the measured total antenna efficiencies for the four radiation nulls are 2%, 5%, 3% and 1.5% from the lower to higher frequency. Average efficiencies in two passbands are over 85% and 87%, respectively. Fig. 12 shows the simulated and measured radiation patterns before and after bending. The proposed antenna achieves measured F/B ratios more than 10 dB. It is worth noting that, the F/B ratios can be further improved by increasing the number of directors and adding reflectors. Besides, the measured cross-polarization levels are all less than −20 dB. A good agreement is observed between the simulated and measured results.

From table I, to the best of our knowledge, none of previously reported dual-band endfire antennas has flexible conformability whereas. Besides, with four radiation nulls, the proposed antenna can adjust operating frequencies and out-of-band radiation suppression of dual-band flexibly. Moreover, our antenna has the lowest profile and overall dimension. Meanwhile, this antenna design offers flexible conformability, enabling reliable performance on curved surfaces, along with independently tunable dual-band operation.

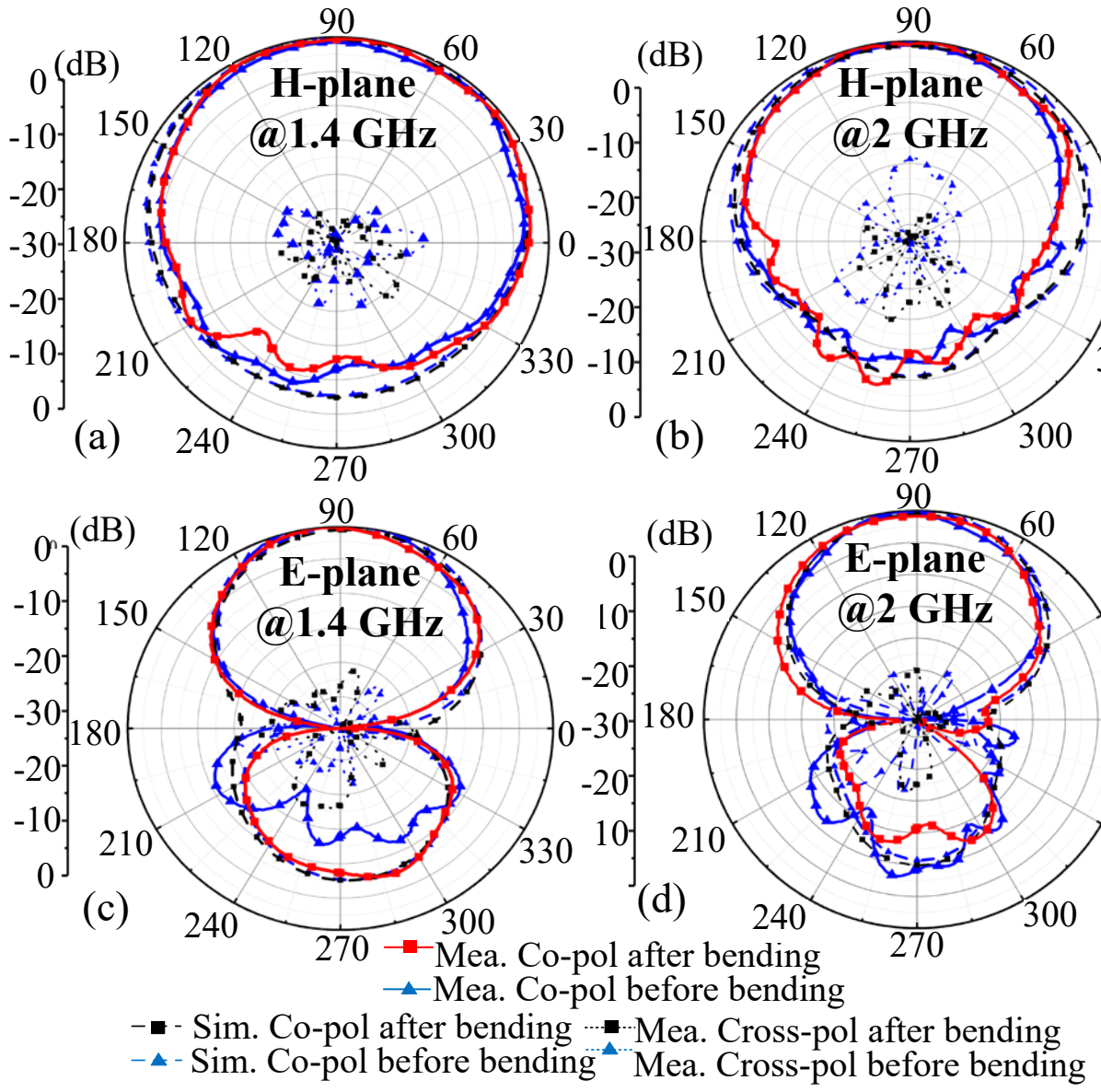

**Fig. 12.** Simulated and measured radiation patterns of the proposed antenna.

TABLE I
Comparison Of The Work with Related Works

| Ref. | Dimension ($\lambda_0^3$) | Suppression (dB) | Flexible Conformal | Nulls | Fre. Select (GHz) | F/B (dB) |
|---|---|---|---|---|---|---|
| [15] | 0.45×0.45 ×0.01 | > 11 | No | 3 | 3.5/4.6 | 15 |
| [19] | 0.41×0.30 ×0.015 | / | No | / | 4.2/7.1 | 15 |
| [20] | 0.29×0.36 ×0.004 | / | No | / | 1.6/2.6 | 9.4 |
| [21] | 0.96×1.09 ×0.032 | / | No | / | 9.6/11 | 10 |
| **This Work** | **0.56×0.56 ×0.0004** | **> 11** | **Yes** | **4** | **1.4/2** | **10** |

## IV. Conclusion

In this letter, a single-layer dual-band flexible conformal EFA without extra circuit is presented. The dual-band endfire radiation is realized by incorporating reflecting structures into co-designed dual-band dipoles. The filtering functionality is achieved through a combination of electric and magnetic coupling mechanisms. Independent tunability of the two frequency bands is enabled by the frequency-selective open-circuit characteristics of dual-band EFA structure. Moreover, the electromagnetic performance remains stable following flexible conformal adaptation. These attributes make the design well-suited for emergency communication applications.